\documentclass[11pt]{article}

\usepackage{geometry}
\usepackage{amsmath,amssymb}
\usepackage{natbib}
\usepackage{booktabs}
\usepackage{graphicx}
\usepackage[colorlinks,citecolor=green,linkcolor=blue,bookmarks=false,hypertexnames=true]{hyperref}

\renewcommand{\cite}{\citep}

\renewcommand{\toprule}{%
  \noalign{\vskip\abovetopsep}%
  \hline\hline
  \noalign{\vskip\belowrulesep}%
}
\renewcommand{\midrule}{%
  \noalign{\vskip\aboverulesep}%
  \hline
  \noalign{\vskip\belowrulesep}%
}
\renewcommand{\bottomrule}{%
  \noalign{\vskip\aboverulesep}%
  \hline
  \noalign{\vskip\belowbottomsep}%
}

\usepackage{titling}
\usepackage{array}
\preauthor{\begin{center}
    \large \lineskip .75em%
        \begin{tabular}[t]{>{\centering\arraybackslash}p{.45\textwidth}}}
        \postauthor{\end{tabular}\par\end{center}}
\makeatletter
\renewcommand\and{
  \end{tabular}%
  \hfill
  \begin{tabular}[t]{>{\centering\arraybackslash}p{.45\textwidth}}}
\makeatother

\title{The material life cycle of \emph{Koobor}: separating coherent organization from potential-vorticity evolution in a wildfire-generated stratospheric vortex}

\author{F.\ Andrade-Canto\thanks{Corresponding author.}\\ Departamento de Observaci\'on y Estudio de la Tierra, la Atm\'osfera y el Oc\'eano\\ El Colegio de la Frontera Sur\\ Chetumal, Quintana Roo, Mexico\\ \href{mailto:fberon@miami.edu}{\texttt{fernando.andrade@ecosur.mx}}\and \and F.J.\ Beron-Vera\\ Department of Atmospheric Sciences\\ Rosenstiel School of Marine, Atmospheric \& Earth Science\\ University of Miami\\ Miami, Florida, USA\\ \href{mailto:fberon@miami.edu}{\texttt{fberon@miami.edu}}}

\date{Started: October 31, 2025.  This version: \today.}

\begin{document}

\maketitle

\begin{abstract}
Pyro-cumulonimbus convection associated with extreme wildfires can generate long-lived stratospheric vortices, yet their material coherence and life cycle remain poorly characterized. We apply geodesic vortex detection to winds from reanalysis to identify objectively defined (observer-independent) material boundaries of \emph{Koobor}, generated by the 2019--2020 Australian bushfires. These boundaries are closed material curves that undergo uniform finite-time tangential stretching and resist filamentation. We reconstruct their evolution across isentropic levels.

Material coherence develops upward from 590~K, reaches maximum persistence at 690~K, and becomes progressively shorter-lived at higher levels. At 690~K, the reconstructed birth-to-death interval is approximately 40~days, while overlapping detections across levels define a nearly 60-day envelope. Analysis of potential vorticity (PV) within and around these material boundaries reveals a distinct dynamical evolution: the material boundary forms before strong PV isolation develops, a compact PV core emerges during the mature phase, and PV isolation weakens as material coherence approaches its demise. Strong PV isolation is concentrated at intermediate levels, whereas at 850~K material coherence occurs without a comparably well-defined PV core.

These results provide an objective material description of the birth, vertical development, and decay of a wildfire-generated stratospheric vortex and show that material organization precedes the development of its strongest PV signature, which subsequently intensifies and weakens over the lifetime of the coherent vortex.
\end{abstract}

\section*{Plain Language Summary}

Extreme wildfires can inject smoke into the stratosphere, where solar heating of the smoke can produce long-lived rotating structures. We study \emph{Koobor}, the prominent vortex generated by the 2019--2020 Australian bushfires, using a method that identifies material boundaries that resist deformation and filamentation. The coherent vortex develops upward through the stratosphere, persists longest near 690~K, and disappears earlier at higher levels. Together, detections at different levels describe a life cycle spanning nearly 60~days. We also examine potential vorticity within and around these material boundaries. Strong potential-vorticity isolation develops after the materially coherent core has formed and weakens as material coherence approaches its end, showing that the strong potential-vorticity signature emerges during, rather than defines, the life cycle of the coherent vortex.

\section*{Key Points}

\begin{itemize}
    \item Geodesic vortex detection identifies finite-time material boundaries of the wildfire-generated stratospheric vortex \emph{Koobor} that resist filamentation.
    \item Material coherence develops upward, persists longest at 690~K, and becomes progressively shorter-lived at higher isentropic levels.
    \item Strong PV isolation develops after material coherence is established and weakens toward its demise, while remaining comparatively weak at 850~K.
\end{itemize}

\section{Introduction}

Extreme wildfires can generate pyro-cumulonimbus (pyroCb) convection capable of injecting smoke and other combustion products directly into the lower stratosphere \citep{Fromm-etal-10,Peterson-etal-18,Yu-etal-19}. Absorption of solar radiation by black carbon produces diabatic heating and self-lofting of the injected plume, allowing smoke to ascend several kilometers after injection and remain in the stratosphere for months \citep{deLaat-etal-12,Yu-etal-19}. Under sufficiently intense forcing, this process can generate compact, long-lived anticyclonic vortices in which smoke, thermal anomalies, and the circulation evolve together.

The 2019--2020 Australian wildfires produced an exceptional example of this phenomenon \citep{Khaykin-etal-20,Kablick-etal-20,Allen-etal-20}. The most prominent smoke-charged vortex, subsequently named \emph{Koobor}, persisted for several weeks while circling the Southern Hemisphere and ascending to approximately 35~km. Observations and analyses revealed a compact anticyclonic circulation accompanied by strong aerosol, temperature, ozone, and potential-vorticity anomalies, with radiative heating of the concentrated smoke driving continued diabatic ascent \citep{Khaykin-etal-20,Kablick-etal-20,Allen-etal-20,Lestrelin-etal-21}. Subsequent theoretical and numerical work has clarified how localized diabatic heating by an ascending smoke plume can generate and sustain an anticyclonic potential-vorticity anomaly \citep{Shah-Haynes-24,Podglajen-etal-24}. \emph{Koobor} therefore provides an unusually well-observed example of a vortex generated and sustained by diabatic forcing rather than simply advected as a passive atmospheric feature.

Despite this detailed characterization, a basic dynamical question remains: to what extent was \emph{Koobor} a materially coherent vortex? Potential vorticity (PV) provides a fundamental dynamical description of stratospheric vortices, but a compact PV anomaly or a sharp PV gradient does not by itself identify the boundary of a materially coherent air mass. More generally, the relation between PV structure and Lagrangian transport barriers is not one-to-one. Early work on the stratospheric polar vortex related its material isolation to robust Lagrangian barriers near the polar jet \citep{Rypina-etal-07a}, while subsequent studies showed that Lagrangian transport barriers need not coincide with PV barriers, first in idealized PV-conserving flows and later in realistic stratospheric flows \citep{Beron-etal-08-JAS,Beron-etal-12}. Geodesic vortex detection further showed that PV-based estimates of the stratospheric polar-vortex edge can either underestimate or overestimate the outermost materially coherent boundary \citep{Serra-etal-17}. Material coherence instead concerns the deformation and transport of fluid parcels and requires identification of material boundaries that resist filamentation over finite times. This distinction is particularly relevant for a diabatically evolving smoke vortex, whose thermodynamic and PV signatures may change substantially during its lifetime.

Previous studies have characterized different aspects of \emph{Koobor}. Relative vorticity, PV, temperature, ozone, and aerosol observations have documented its atmospheric structure, ascent, and persistence \citep{Khaykin-etal-20,Lestrelin-etal-21}. \citet{Curbelo-Rypina-23} analyzed the three-dimensional organization of smoke transport using finite-time Lagrangian diagnostics. More recently, an undergraduate thesis applied the Lagrangian-averaged vorticity deviation (LAVD) criterion for rotational coherence \citep{Haller-etal-16} to the Australian event at several isentropic levels \citep{Llorach-24}. These studies establish the remarkable dynamical and thermodynamical evolution of \emph{Koobor}, but leave open a distinct question: when did its materially coherent core form, how long did it persist, and when did it lose coherence?

We address this question using geodesic vortex detection \citep{Haller-Beron-13,Haller-Beron-14,Haller-15,Andrade-etal-25}. The method identifies objectively defined (observer-independent) vortex boundaries as closed material curves exhibiting uniform finite-time tangential stretching and hence strong resistance to filamentation. By systematically varying the initialization time and coherence interval \citep{Andrade-etal-20}, we reconstruct the birth, persistence, and decay of these material boundaries at successive isentropic levels. The resulting life cycle is level-resolved: each detected boundary is material over its corresponding finite-time interval, while boundaries identified at different levels are not assumed to constitute a single three-dimensional material surface.

Applying this framework to reanalyzed winds, we reconstruct the birth, vertical development, and decay of the materially coherent core of \emph{Koobor}. We show that material coherence develops upward through the stratosphere, reaches its greatest persistence at an intermediate isentropic level, and becomes markedly shorter-lived aloft. We then examine the evolving PV structure within and around the objectively defined material boundaries. The PV analysis is not used to define the vortex edge; rather, it reveals how the dynamical signature of the vortex develops over the reconstructed material life cycle. In particular, strong PV isolation develops after the material boundary forms and weakens as material coherence approaches its demise. Thus, geodesic detection determines the existence and lifetime of the coherent material core, while the PV signature emerges and evolves within that material organization.

The rest of the paper is organized as follows. Section~\ref{sec:methods} describes the geodesic vortex-detection framework, the life-cycle reconstruction procedure, the isentropic approximation, the reanalysis data and numerical implementation, and the potential-vorticity analysis. Section~\ref{sec:results} presents the reconstructed material life cycle of \emph{Koobor}, including its vertical development, persistence, and decay across isentropic levels, together with the associated potential-vorticity evolution. Section~\ref{sec:discussion} discusses the implications of the results in the context of diabatically generated stratospheric vortices and previous studies of \emph{Koobor}. Section~\ref{sec:conclusions} summarizes the main findings.

\section{Methodology}\label{sec:methods}

\subsection{Material vortex boundaries}

We analyze atmospheric motion on isentropic surfaces using a finite-time Lagrangian framework. For a fixed potential-temperature level $\theta$, let $M_\theta$ denote the corresponding two-dimensional isentropic surface and $D\subset M_\theta$ the flow domain, with $\mathbf{x}$ denoting coordinates on $M_\theta$. The motion of air parcels tangent to $M_\theta$ is described by
\begin{equation}
    \dot{\mathbf{x}}=\mathbf{u}(\mathbf{x},t),\qquad
    \mathbf{x}(t_0)=\mathbf{x}_0\in D,
\end{equation}
with associated flow map
\begin{equation}
    F_{t_0}^{t}:D\rightarrow D.
\end{equation}

Material deformation over the finite interval $[t_0,t_0+T]$ is quantified by the right Cauchy--Green strain tensor. Because the motion is evaluated on a curved isentropic surface, we use the surface-metric formulation of \citet{Andrade-etal-25}. In coordinates on $M_\theta$, the Cauchy--Green tensor is represented by
\begin{equation}
    C_{t_0}^{t_0+T}(\mathbf{x}_0)
    =
    \nabla F_{t_0}^{t_0+T}(\mathbf{x}_0)^\top
    G\!\left(F_{t_0}^{t_0+T}(\mathbf{x}_0)\right)
    \nabla F_{t_0}^{t_0+T}(\mathbf{x}_0),
\end{equation}
where $G(\mathbf{x})$ represents, in the chosen coordinates, the Riemannian metric on $M_\theta$ inherited from the Euclidean geometry of the surrounding three-dimensional space. The relevant eigenvalues and eigenvectors are defined by the generalized eigenvalue problem
\begin{equation}
    C_{t_0}^{t_0+T}(\mathbf{x}_0)\boldsymbol{\xi}_i(\mathbf{x}_0)
    =
    \lambda_i(\mathbf{x}_0)
    G(\mathbf{x}_0)\boldsymbol{\xi}_i(\mathbf{x}_0),
    \qquad i=1,2,
\end{equation}
with
\begin{equation}
    0<\lambda_1(\mathbf{x}_0)\leq\lambda_2(\mathbf{x}_0),
    \qquad
    \boldsymbol{\xi}_i(\mathbf{x}_0)^\top
    G(\mathbf{x}_0)
    \boldsymbol{\xi}_j(\mathbf{x}_0)
    =\delta_{ij}.
\end{equation}

We identify materially coherent vortex boundaries using the geodesic theory of elliptic Lagrangian coherent structures \citep{Haller-Beron-13,Haller-Beron-14,Haller-15}. These boundaries are stationary curves of an averaged finite-time stretching functional and therefore undergo spatially uniform tangential stretching. Equivalently, they are closed curves $\mathbf{x}_0(s)$ satisfying
\begin{equation}
    \mathbf{x}_0'(s)
    =
    \sqrt{\frac{\lambda_2(\mathbf{x}_0)-p^2}
    {\lambda_2(\mathbf{x}_0)-\lambda_1(\mathbf{x}_0)}}\,
    \boldsymbol{\xi}_1(\mathbf{x}_0)
    \pm
    \sqrt{\frac{p^2-\lambda_1(\mathbf{x}_0)}
    {\lambda_2(\mathbf{x}_0)-\lambda_1(\mathbf{x}_0)}}\,
    \boldsymbol{\xi}_2(\mathbf{x}_0),
\end{equation}
where $\lambda_1(\mathbf{x}_0)<p^2<\lambda_2(\mathbf{x}_0)$ and $p$ is the spatially uniform finite-time tangential stretching factor. Among nested closed $p$-loops, the outermost loop defines the material vortex boundary.

The defining property of these curves is not zero deformation, nor is $p$ required to equal unity. Rather, every material element of the boundary experiences the same finite-time tangential stretching factor. The boundary therefore avoids the differential stretching responsible for filamentation and provides an objective, observer-independent definition of finite-time material coherence. Closed $p$-loops may equivalently be interpreted as null-geodesics of a generalized Green--Lagrange strain tensor \citep{Haller-Beron-13,Haller-Beron-14}.

\subsection{Life-cycle reconstruction}

A material vortex boundary is defined relative to a finite interval $[t_0,t_0+T]$. To determine the life cycle of the coherent vortex rather than its state at a single initialization time, we systematically vary both $t_0$ and $T$, following the approach introduced by \citet{Andrade-etal-20}.

For each sampled initialization time $t_0$, we increase the integration interval while a closed geodesic vortex boundary can be detected. We denote the largest successful forward- and backward-time intervals by
\begin{equation}
    T_{\max}^{+}(t_0)
    \qquad\mbox{and}\qquad
    T_{\max}^{-}(t_0),
\end{equation}
respectively. Each detection therefore supplies two possible temporal endpoints,
\begin{equation}
    t_0+T_{\max}^{+}(t_0),
    \qquad
    t_0-T_{\max}^{-}(t_0).
\end{equation}
The reconstructed birth and death times at a given isentropic level are defined by
\begin{equation}
    t_{\rm birth}
    =
    \min_{t_0}
    \left[t_0-T_{\max}^{-}(t_0)\right],
    \qquad
    t_{\rm death}
    =
    \max_{t_0}
    \left[t_0+T_{\max}^{+}(t_0)\right].
\end{equation}

This construction distinguishes the coherence interval of an individual material boundary from the life cycle of the coherent-vortex state. The latter can exceed any individual $T_{\max}$ because it is reconstructed from a sequence of overlapping finite-time detections. Initialization times are sampled every 5~days, except at the highest analyzed level, where a 3-day spacing is used because the detected intervals are shorter. We impose an upper numerical search horizon of $\tau=40$~days; $\tau$ only bounds the intervals explored and does not define the inferred vortex lifetime.

The procedure is applied independently at each isentropic level. Consequently, the earliest birth and latest death across levels define a level-resolved temporal envelope, not the coherence interval of a single material boundary. Likewise, boundaries detected simultaneously at different isentropic levels are not interpreted as sections of a single three-dimensional material surface.

\subsection{Isentropic approximation}

The geodesic vortex-detection framework used here is two-dimensional and is therefore applied independently on surfaces of constant potential temperature. This formulation is appropriate for diagnosing finite-time horizontal material organization in a strongly stratified atmosphere, but diabatic motion associated with the smoke plume permits parcels to cross isentropic surfaces. The detected boundaries therefore establish material coherence tangent to a selected isentropic surface over their corresponding finite-time intervals; they do not constrain cross-isentropic transport or define a three-dimensional material vortex.

This distinction is especially important for \emph{Koobor}, whose smoke-induced radiative heating produced substantial diabatic ascent. Rather than interpreting detections at different levels as a vertically connected material surface, we use them to determine how the occurrence and persistence of horizontal material coherence vary with potential temperature. The resulting level-resolved reconstruction provides a description of the upward development and eventual disappearance of the coherent core.

\subsection{Data and numerical implementation}

We use winds from the European Centre for Medium-Range Weather Forecasts (ECMWF) ERA5 reanalysis \citep{Hersbach-etal-20} on the 550, 590, 630, 690, 770, 850, and 964~K isentropic surfaces during the 2019--2020 Australian wildfire event. The winds are extracted on isentropic surfaces as provided by ERA5; no vertical interpolation to isentropic levels is performed as part of our analysis. The velocity fields have a horizontal resolution of approximately 31~km and are sampled every 6~h.

Trajectory integration and geodesic vortex detection are performed with the curved-surface implementation of the Julia package CoherentStructures.jl described by \citet{Andrade-etal-25}. Velocities are interpolated cubically in space and time, and trajectories are integrated with a fifth-order Runge--Kutta scheme \citep{Tsitouras-11}. On each isentropic surface, the Cauchy--Green tensor is evaluated on a $256\times256$ grid over a computational domain selected to contain the smoke signature reported by \citet{Khaykin-etal-20}. Closed $p$-loops are obtained by integrating the corresponding line field and identifying limit cycles with a Poincar\'e return map; among nested admissible loops, the outermost is retained as the vortex boundary.

\subsection{Potential-vorticity analysis}

To place the objectively determined material life cycle in its atmospheric-dynamical context, we examine ERA5 potential vorticity on the same isentropic surfaces and times used for the geodesic analysis. As for the winds, potential vorticity is extracted on isentropic surfaces as provided by ERA5. Potential vorticity is used here as a dynamical diagnostic, not as a criterion for defining the vortex boundary. We examine its evolving structure within and around the observer-independent geodesic boundaries during the formation, mature stage, and decay of the materially coherent core, and across the isentropic levels over which coherence is detected.

\section{Results}\label{sec:results}

Geodesic vortex detection reveals a materially coherent core of \emph{Koobor} over a broad range of isentropic levels, with a pronounced level dependence in its formation and persistence. No coherent boundaries are detected at 550 or 964~K, whereas detections between 590 and 850~K describe an upward-developing and subsequently decaying coherent structure. We first examine the life cycle at 690~K, where material coherence is most persistent, and then synthesize the detections across levels.

\subsection{Life cycle at 690~K}

\begin{figure}[t]
    \centering
    \includegraphics[width=\textwidth]{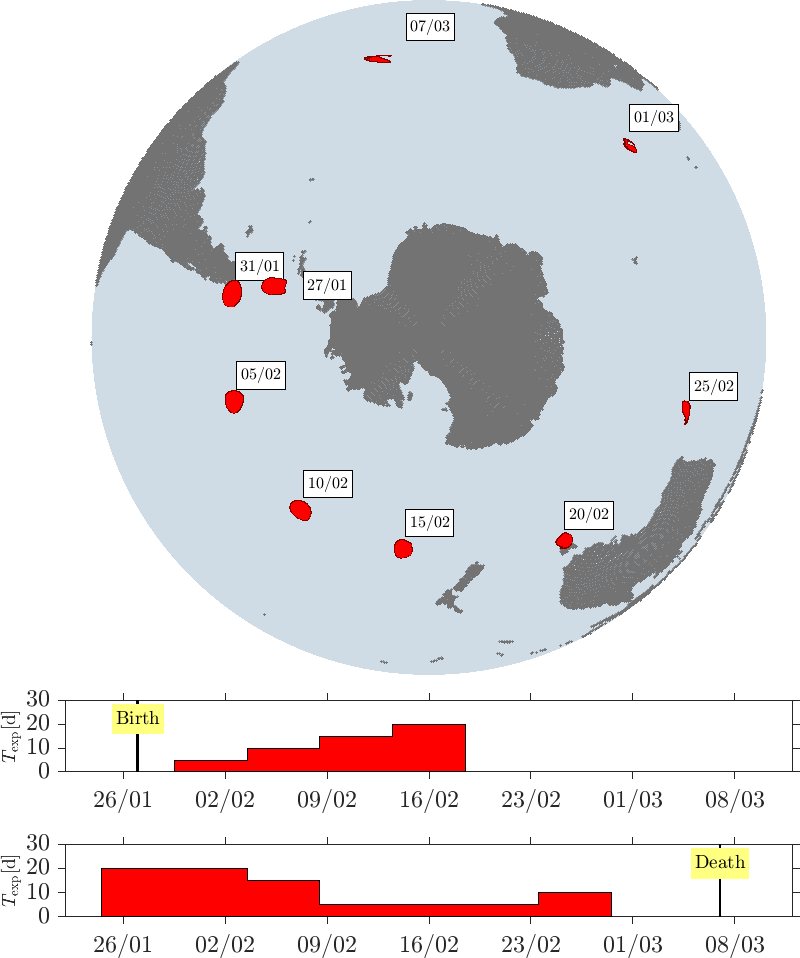}
    \caption{Geodesic vortex detection at the 690-K isentropic level. Top: selected coherent material boundaries along the trajectory of \emph{Koobor}. Bottom: maximum coherence intervals $T_{\rm exp}$ obtained from forward- and backward-time analyses as a function of initialization time. The inferred birth and death dates are indicated.}
    \label{fig:life690}
\end{figure}

Figure~\ref{fig:life690} shows the geodesically detected boundaries at 690~K together with the maximum finite-time coherence intervals obtained from forward- and backward-time analyses. The detected vortex travels around a substantial portion of the Southern Hemisphere, from the region between Patagonia and Antarctica toward southern Africa, while remaining identifiable through successive materially coherent boundaries.

The backward-time analysis places the birth of the coherent vortex on 27 January 2020. Forward-time detections initialized after this date persist for progressively different intervals, reaching approximately 20~days for initialization times in mid-February. The corresponding forward-time analysis places the death on 7 March 2020. Thus, at 690~K, the reconstructed birth-to-death interval is approximately 40~days.

This interval is distinct from the coherence time of any individual material boundary. Each boundary in Fig.~\ref{fig:life690} is materially coherent only over the finite interval associated with its particular initialization time. The approximately 40-day life cycle instead results from the temporal succession and overlap of these detections. In this sense, the analysis identifies a persistent \emph{state of material coherence} supported by different material boundaries during different portions of the reconstructed life cycle.

\subsection{Vertical development of material coherence}

\begin{figure}[t]
    \centering
    \includegraphics[width=\textwidth]{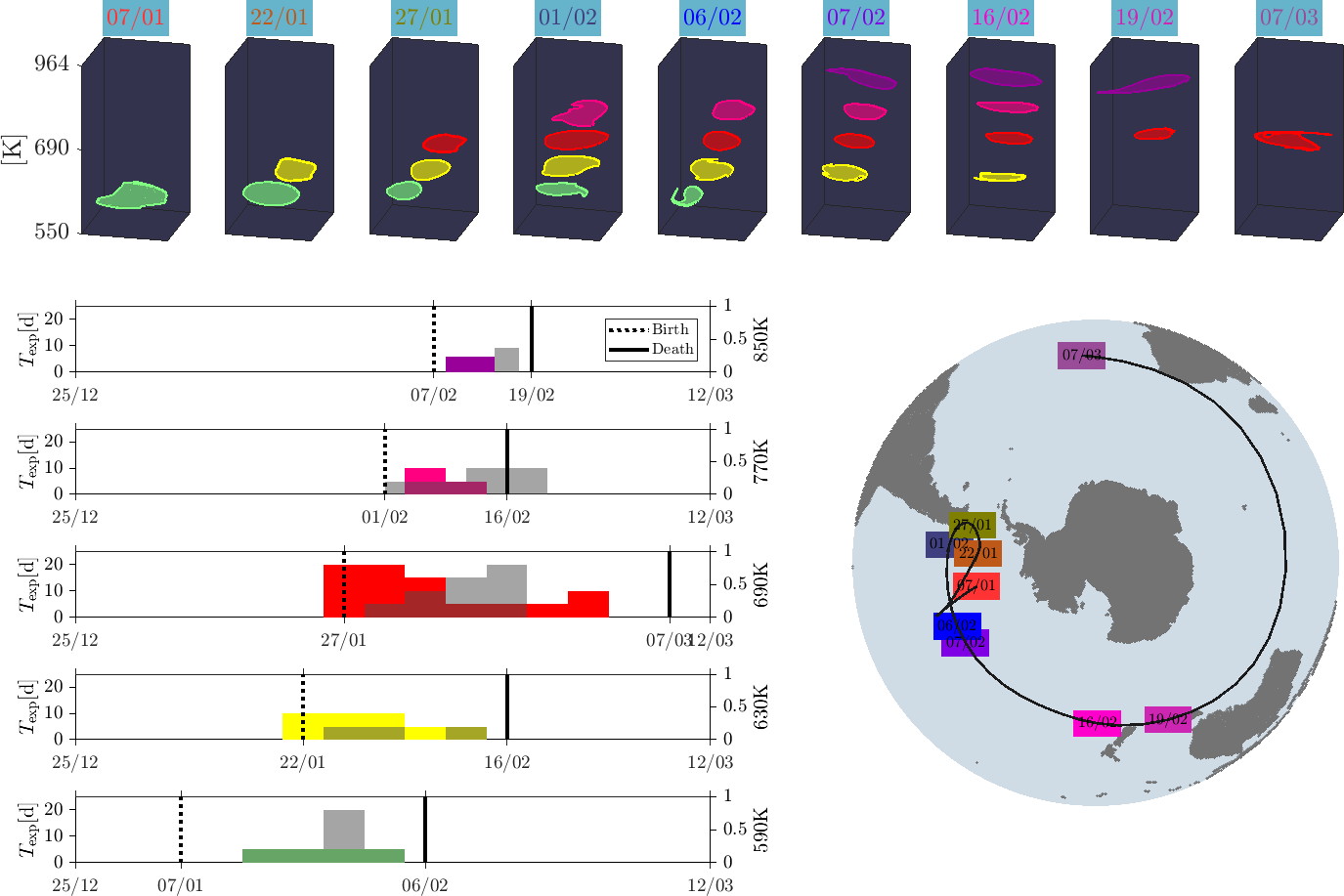}
    \caption{Level-resolved evolution of material coherence in \emph{Koobor}. Top: geodesically detected material boundaries at selected dates and isentropic levels. Bottom left: maximum coherence intervals as functions of initialization time at 850, 770, 690, 630, and 590~K, with inferred birth and death dates indicated. Bottom right: trajectory of the coherent vortex during the analyzed period. Colors identify the corresponding isentropic levels.}
    \label{fig:vertical}
\end{figure}

Repeating the analysis across isentropic levels reveals a clear vertical organization of the coherent vortex (Fig.~\ref{fig:vertical}). Material coherence first appears at 590~K on 7 January 2020, followed by 630~K on 22 January and 690~K on 27 January. Coherent boundaries appear later at the upper levels, on 6 February at 770~K and 7 February at 850~K. The onset of material coherence therefore progresses upward through the analyzed isentropic range.

The persistence of the coherent vortex is strongly nonmonotonic with potential temperature. The reconstructed birth-to-death intervals are approximately 30~days at 590~K, 25~days at 630~K, 40~days at 690~K, 15~days at 770~K, and 12~days at 850~K (Table~\ref{tab:lifecycle}). Material coherence is therefore longest lived at the intermediate 690-K level and becomes markedly shorter-lived above it. The upper-level coherent boundaries disappear by 19 February at 850~K and 21 February at 770~K, while coherence at 690~K persists into early March.

\begin{table}[t]
\centering
\begin{tabular}{lcccc}
\toprule
Level & Birth & Death & Life-cycle interval & Radius \\
\midrule
590~K & 7 Jan 2020  & 6 Feb 2020  & 30~d & 104~km \\
630~K & 22 Jan 2020 & 16 Feb 2020 & 25~d & 106~km \\
690~K & 27 Jan 2020 & 7 Mar 2020  & 40~d & 111~km \\
770~K & 6 Feb 2020  & 21 Feb 2020 & 15~d & 118~km \\
850~K & 7 Feb 2020  & 19 Feb 2020 & 12~d & 138~km \\
\bottomrule
\end{tabular}
\caption{Level-resolved life-cycle properties of \emph{Koobor}.}
\label{tab:lifecycle}
\end{table}

Figure~\ref{fig:vertical} summarizes this evolution both spatially and temporally. The sequence of level-resolved boundaries shows the coherent structure developing upward during January and reaching its greatest vertical extent in early February. Thereafter, coherence is lost first at the highest levels and subsequently at lower levels, while the 690-K core remains detectable longest. The simultaneous occurrence of boundaries on several isentropic surfaces during early February demonstrates vertical organization of the material-coherence pattern, without implying that these independently detected curves form a single three-dimensional material surface.

Taken together, the level-resolved detections span nearly 60~days, from the first coherent appearance at 590~K on 7 January to the latest reconstructed decay at 690~K on 7 March. This interval is a temporal envelope of the vertically evolving coherent-vortex state, assembled from overlapping finite-time detections at different levels, rather than the coherence time of a single material boundary.

\subsection{Evolution of the coherent material boundary}

\begin{figure}[t]
    \centering
    \includegraphics[width=\textwidth]{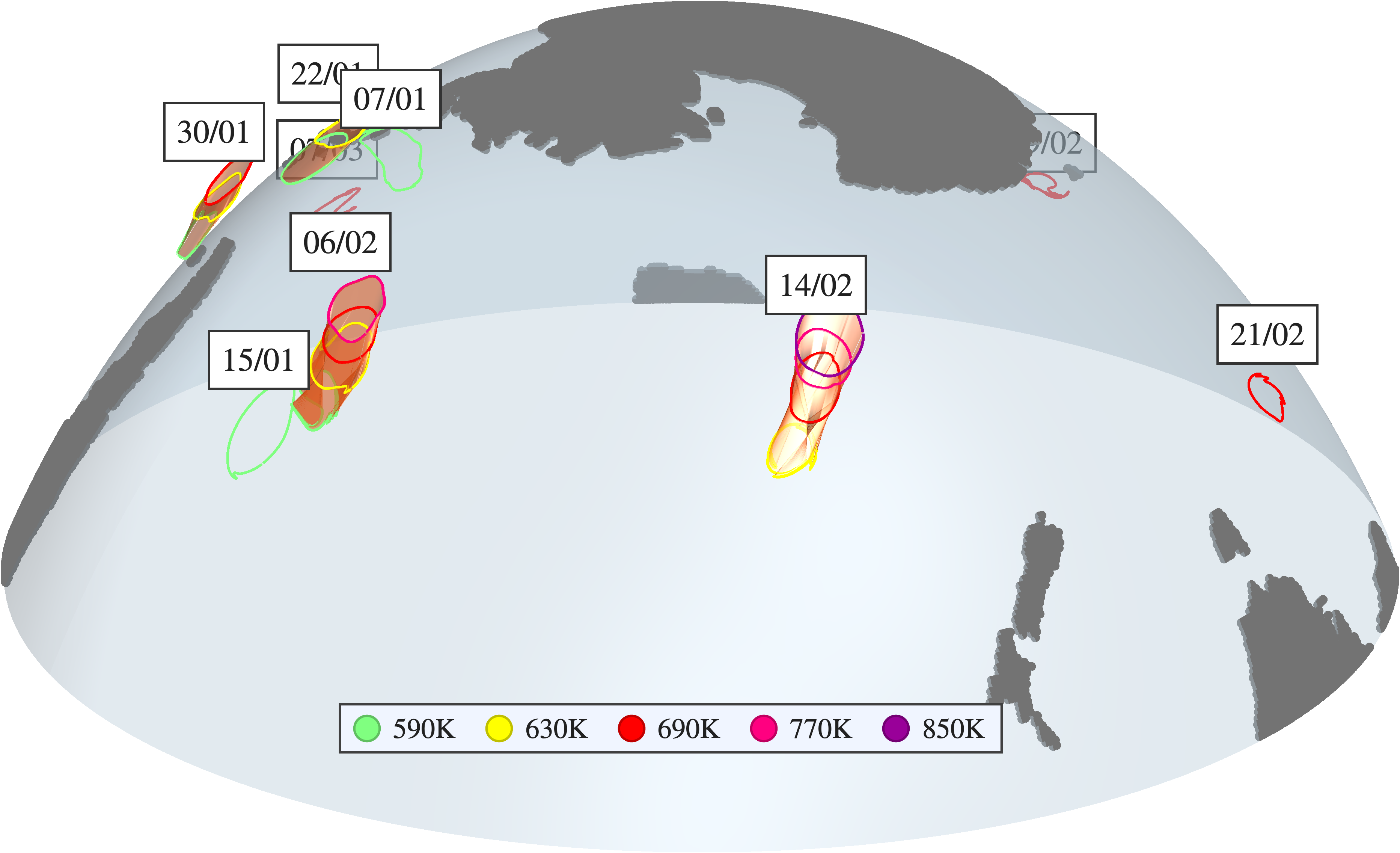}
    \caption{Evolution of the geodesically detected material boundary of \emph{Koobor} at 690~K. Top: selected boundaries along the vortex trajectory between late January and early March 2020. Bottom: forward- and backward-time maximum coherence intervals used to frame the birth and death of the coherent vortex.}
    \label{fig:boundaries690}
\end{figure}

The sequence in Fig.~\ref{fig:boundaries690} illustrates the changing geometry of the coherent core at 690~K. Near its inferred birth, the detected boundary is compact and located close to the Antarctic Peninsula. During February it remains a closed material loop while being transported eastward around the Southern Hemisphere. Its size and shape evolve substantially along the trajectory, with increasingly elongated configurations appearing toward the end of the reconstructed life cycle.

The coherence-time diagrams in Fig.~\ref{fig:boundaries690} place these individual boundaries in their finite-time context. Forward-time coherence increases after the inferred birth and reaches its longest intervals for boundaries initialized during February before decreasing as the vortex approaches its demise. The backward-time calculation provides the complementary constraint on formation. Together, these calculations distinguish the persistence of the coherent-vortex state from the finite coherence interval of each individual boundary.

By 7 March, the final detected structures are strongly elongated, marking the end of the interval over which a closed geodesic material boundary can be identified at 690~K. The resulting evolution---from the appearance of a compact coherent core, through sustained finite-time material coherence, to eventual loss of a closed coherent boundary---provides the material definition of the birth and decay of \emph{Koobor} used here.

\subsection{Potential-vorticity evolution of the materially coherent vortex}

We next examine how the PV structure evolves relative to the material life cycle identified above. Because the large-scale meridional variation of PV changes substantially along the trajectory of \emph{Koobor}, we consider the anomaly
\begin{equation}
q'(\lambda,\vartheta,t)=q(\lambda,\vartheta,t)-\langle q\rangle_{\lambda}(\vartheta,t),
\end{equation}
where $\langle q\rangle_{\lambda}$ denotes the zonal mean on the corresponding isentropic surface. This subtraction is used only to expose the localized PV structure associated with the vortex; it plays no role in determining the geodesic boundary.

For each detected boundary, we compute the mean PV anomaly inside the materially coherent region, $\langle q'\rangle_{\rm in}$, its spatial standard deviation, $\sigma(q')_{\rm in}$, and the interior--exterior contrast
\begin{equation}
\Delta q'=\langle q'\rangle_{\rm in}-\langle q'\rangle_{\rm out},
\end{equation}
where $\langle q'\rangle_{\rm out}$ denotes the corresponding mean outside the material boundary. These quantities characterize, respectively, the strength and internal variability of the PV anomaly and the degree of PV isolation across the material boundary.

\begin{figure}[t]
\centering
\includegraphics[width=\textwidth]{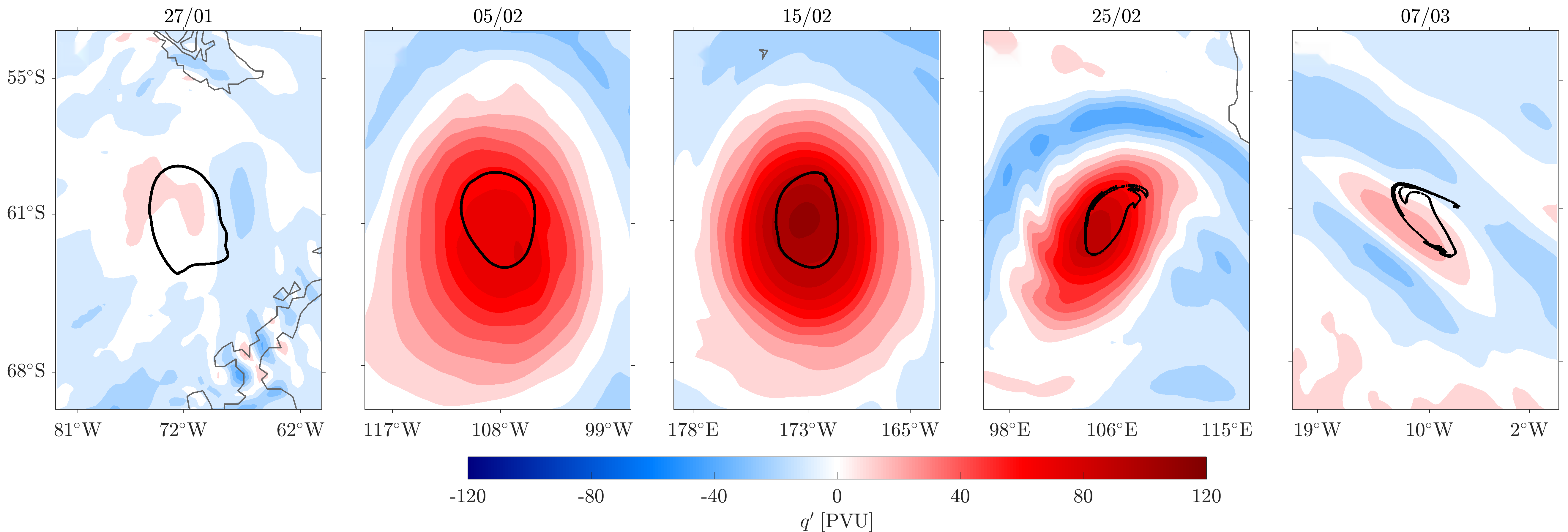}
\caption{Evolution of the potential-vorticity anomaly $q'$ at 690~K during the reconstructed life cycle of the materially coherent \emph{Koobor} vortex. The geodesic material boundary is overlaid where detected. The sequence shows comparatively weak PV isolation near material birth, development of a compact and strongly isolated PV core during the mature phase, and weakening of the PV isolation toward the end of material coherence.}
\label{fig:pv690}
\end{figure}

The evolution at 690~K illustrates a clear formation--maturity--decay sequence (Fig.~\ref{fig:pv690}). At the reconstructed material birth on 27 January, a closed geodesic boundary is already present, while PV isolation is comparatively weak: $\langle q'\rangle_{\rm in}=6.80$~PVU and $\Delta q'=3.61$~PVU. The PV anomaly subsequently intensifies within the materially coherent region, producing a compact and strongly isolated core during February. The mean interior anomaly reaches 112.59~PVU on 19 February, while the interior--exterior contrast reaches 19.41~PVU on 18 February. Toward the reconstructed end of material coherence, the PV isolation weakens substantially. By 7 March, $\langle q'\rangle_{\rm in}$ has decreased to 17.26~PVU and $\Delta q'$ to 1.51~PVU. Thus, the material boundary appears before strong PV isolation develops; the latter emerges during the mature stage and weakens as the materially coherent vortex approaches its demise.

\begin{figure}[t]
\centering
\includegraphics[width=\textwidth]{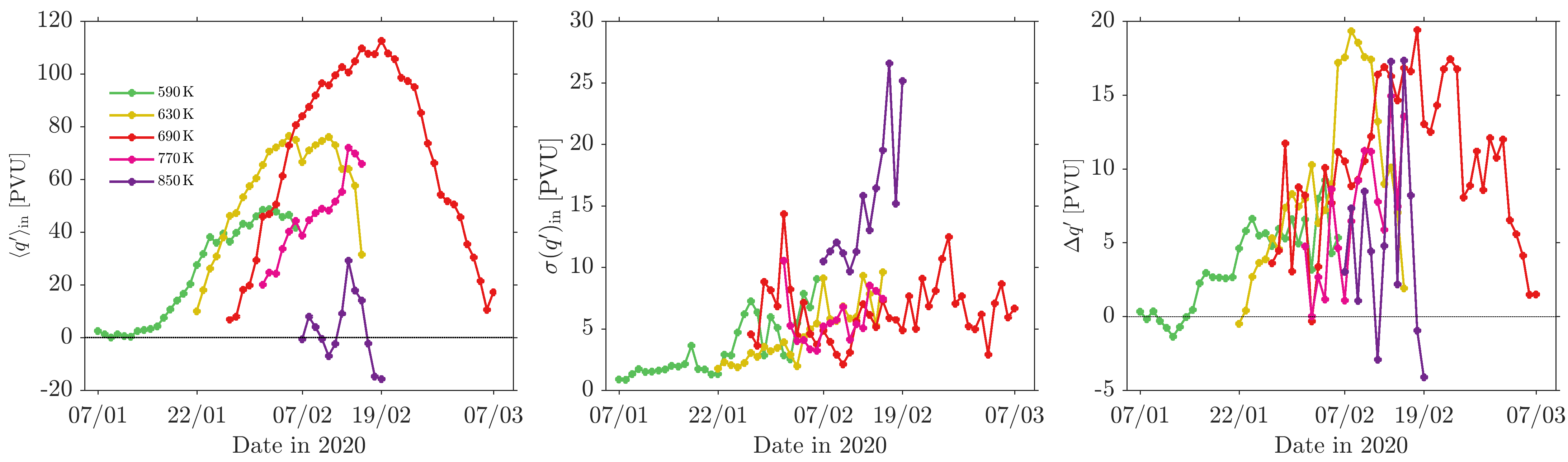}
\caption{Evolution of potential-vorticity diagnostics within and around the geodesic material boundaries at the isentropic levels where material coherence is detected: mean interior anomaly $\langle q'\rangle_{\rm in}$, interior spatial variability $\sigma(q')_{\rm in}$, and interior--exterior contrast $\Delta q'$. Strong and sustained PV isolation is concentrated at intermediate levels, particularly 630--690~K, whereas the upper-level materially coherent regions exhibit weaker and less persistent PV signatures.}
\label{fig:pvlevels}
\end{figure}

The relation between material coherence and PV structure also varies with altitude (Fig.~\ref{fig:pvlevels}). The strongest and most sustained PV isolation occurs at the intermediate 630- and 690-K levels, where positive interior anomalies are accompanied by substantial interior--exterior PV contrasts. At 590~K the PV anomaly develops more gradually, while at higher levels the PV structure becomes less persistent and more variable as the intervals of material coherence shorten. The 850-K level provides the clearest contrast: although a materially coherent boundary is detected, the enclosed PV anomaly is weak and highly heterogeneous, with a mean $\langle q'\rangle_{\rm in}=2.97$~PVU and a mean interior variability $\sigma(q')_{\rm in}=15.20$~PVU over the detected interval. The interior--exterior contrast is also highly variable and changes sign on some dates. Material coherence at this level therefore occurs without the compact, persistently isolated PV core seen at 630--690~K.

These results separate two features of the evolution of \emph{Koobor} that need not coincide. The geodesic boundaries identify the finite-time material organization of the vortex, whereas the strong PV anomaly develops within that organization after its formation. At 690~K the PV core intensifies markedly during the material lifetime and weakens toward its end, while at 850~K a material boundary exists without a comparably well-defined PV core. A strong PV anomaly is therefore a prominent signature of the mature state of \emph{Koobor}, but is not a prerequisite for the appearance of its materially coherent boundary.

\section{Discussion}\label{sec:discussion}

The reconstructed material life cycle and the accompanying PV evolution provide two pieces of information that are closely related but not interchangeable. The geodesic calculation identifies when a coherent air mass exists from the finite-time deformation induced by the velocity field. That same flow advects and deforms PV, while diabatic processes can additionally generate or modify PV and the resulting PV distribution can feed back dynamically on the circulation. At 690~K, the materially coherent boundary is already present when PV isolation is comparatively weak; the PV anomaly subsequently intensifies into a compact core and weakens again as material coherence approaches its demise. The existence of material coherence without a comparably well-defined PV core at 850~K further shows that a strong PV anomaly is not a prerequisite for the detected material organization.

This temporal ordering can be placed in the context of the observed development of \emph{Koobor}. \citet{Khaykin-etal-20} documented the organization of the initially dispersed smoke into a compact bubble and its subsequent association with a persistent anticyclonic circulation, interpreting absorption of solar radiation by black carbon as the source of localized diabatic heating and continued ascent. They further argued that confinement by the vortex was required to preserve the compact smoke structure. Our results add a material perspective to this picture: the finite-time organization of the flow into a coherent region can be identified before the strong PV isolation characteristic of the mature vortex is established.

Idealized dynamical studies provide a possible physical interpretation of this sequence. \citet{Shah-Haynes-24} showed that localized heating associated with an ascending tracer distribution can generate a co-moving anticyclonic PV anomaly, while \citet{Podglajen-etal-24} investigated the dynamics of diabatically forced anticyclonic plumes and the role of the resulting low-PV vortex in isolating the heated plume. Viewed in this context, the strengthening PV anomaly within the materially coherent region may reflect both the organization of PV by the coherent flow and its diabatic modification as the smoke-charged vortex develops. The resulting PV anomaly may, in turn, dynamically reinforce the circulation and its resistance to environmental deformation. Establishing the relative roles and causal feedbacks among material confinement, diabatic PV generation, and the evolving circulation is, however, beyond the kinematic analysis performed here: the geodesic calculation diagnoses material coherence from the reanalyzed velocity field but does not determine the dynamical mechanisms responsible for producing or maintaining that velocity field.

The vertical dependence is consistent with this picture. The strongest and most persistent PV isolation occurs at intermediate isentropic levels, where the material life cycle is also longest, whereas the shorter-lived coherent regions aloft need not contain a similarly compact PV core. Thus, the mature PV anomaly and prolonged material coherence become closely associated during the developed stage of \emph{Koobor}, without implying that the PV anomaly itself defines the material vortex boundary. This distinction is particularly important here because diabatic forcing prevents PV from behaving as a materially conserved tracer: its spatial organization reflects advection and deformation by the flow together with diabatic modification and dynamical feedback.

\section{Conclusions}\label{sec:conclusions}

We have provided a finite-time material characterization of \emph{Koobor}, the long-lived stratospheric vortex generated by the 2019--2020 Australian wildfires. Geodesic vortex detection applied to ERA5 winds identifies objectively defined (observer-independent) material boundaries and reconstructs the formation, persistence, and loss of the coherent core without relying on a prescribed instantaneous vortex signature.

The reconstruction reveals a pronounced vertical organization. Material coherence first appears at 590~K, develops upward, and persists longest at 690~K, where the reconstructed birth-to-death interval is approximately 40~days. It becomes substantially shorter-lived at higher levels, while overlapping detections across levels define a nearly 60-day envelope. These intervals describe an evolving, level-resolved coherent-vortex state rather than the lifetime of a single material boundary or three-dimensional material surface.

The PV evolution reveals a distinct dynamical progression within this material life cycle. At 690~K, the material boundary forms before strong PV isolation develops; a compact PV core subsequently emerges during the mature phase and weakens toward the demise of material coherence. Strong PV isolation is concentrated at intermediate levels, whereas at 850~K material coherence occurs without a comparably well-defined PV core. Thus, the material organization of \emph{Koobor} precedes the development of its strongest PV signature, distinguishing the finite-time coherence of the air mass from the evolving PV structure associated with it.

These results suggest a picture in which material organization, PV evolution, and diabatic forcing are dynamically coupled but need not develop simultaneously. The flow can organize a coherent air mass before a strong PV anomaly emerges, while subsequent advection, diabatic PV modification, and dynamical feedback may shape the mature vortex. Distinguishing these processes provides a basis for understanding how diabatically generated stratospheric vortices acquire, maintain, and ultimately lose material coherence.

\section*{Acknowledgments}

This research received no external funding. No institutional funds were available from ECOSUR (Mexico) or the University of Miami to support publication costs.

\section*{Conflict of Interest}

The authors declare no conflicts of interest.

\section*{Availability Statement}

The atmospheric velocity data are obtained from the European Centre for Medium-Range Weather Forecasts (ECMWF) Reanalysis v5 (ERA5), accessible at \url{https://www.ecmwf.int/en/forecasts/dataset/ecmwf-reanalysis-v5}. These data are publicly available subject to ECMWF terms of use.

The computational framework used in this study builds on the Julia package CoherentStructures.jl, developed by Daniel Karrasch and available at \url{https://github.com/CoherentStructures/CoherentStructures.jl}. An implementation adapted to flows on curved surfaces, developed by Gage Bonner, is available at \url{https://github.com/70Gage70/CoherentLagrangianVortices}.

\bibliographystyle{apalike}
\bibliography{fot}

\end{document}